\documentclass[twocolumn]{jpsj2} 
\title{The Yellow Excitonic Series of Cu$_{2}$O Revisited by Lyman Spectroscopy}

\author{Takeshi \textsc{TAYAGAKI}$^{1}$, Andre \textsc{Mysyrowicz}$^{2}$ and Makoto \textsc{Kuwata-Gonokami}$^{1,3}$\thanks{E-mail address: gonokami@ap.t.u-tokyo.ac.jp}}

\inst{$^{1}$Solution Oriented Research for Science and Technology (SORST), JST, 7-3-1 Hongo, Bunkyo-ku, Tokyo 113-8656, Japan \\
$^{2}$Laboratoire d'Optique Appliquee, ENSTA, Ecole Polytechnique, Palaiseau, France \\
$^{3}$Department of Applied Physics, the University of Tokyo, 7-3-1 Hongo, Bunkyo-ku, Tokyo 113-8656, Japan}

\abst{We report on the observation of the yellow exciton Lyman series up to the fourth term in Cu$_{2}$O by time-resolved mid-infrared spectroscopy. The dependence of the oscillator strength on the principal quantum number $\it{n}$ can be well reproduced using the hydrogenic model including an AC dielectric constant, and precise information on the electronic structure of the 1s exciton state can be obtained. A Bohr radius $a_{1s}$=7.9 $\AA$ and a 1s-2p transition dipole moment $\mu$$_{1s-2p}$= 4.2 e$\AA$ were found. }

\kword{exciton, Lyman series, Cu$_{2}$O, central cell correction, mid-infrared, ultrafast spectroscopy}

\begin{document}
\maketitle

The concept of exciton, a composite quasi-particle in a semiconductor consisting of an electron-hole pair bound by Coulomb attraction, is well established \cite{Wannie37, Mott38, Knox63}. Based on the effective-mass approximation, the internal electronic structure of an exciton can be described by a two-particle Schr\"{o}dinger equation leading to a series of bound states similar to that of atomic hydrogen. In the case of an exciton, however, the Rydberg energy is reduced by several orders of magnitude because of the screening effect of the crystal lattice. It should also be noted that an exciton is the electronic elementary excitation of a system of many interacting electrons, so that a simple two-body picture has fundamental limitations.\cite{Sham66} The many body aspect of excitons has recently attracted renewed interest, related to issues such as the interpretation of excitonic luminescence \cite{Chatte04}, the origin of excitonic nonlinearities \cite{Gonoka97}, and the occurrence of a Bose-Einstein condensation (BEC) phase.

Recent advances in laser technology have made new spectroscopic tools available. In particular, mid, or far-infrared spectroscopy, often called 'THz spectroscopy,' \cite{Kaindl03} allows to probe intraband kinetics of electrons and holes, the collective motion of electron hole ensembles, as well as the transitions between the different terms of an excitonic series. With excitonic Lyman spectroscopy, it is also possible to probe optically inactive excitons such as spin triplets, and invisible excitons lying outside the light cone. \cite{Gonoka04, Kubouc05, Gonoka05}

The yellow excitonic series in Cu$_{2}$O is a textbook example of Wannier excitons. A very clear hydrogenic series is observed up to the 10$^{th}$ term in linear absorption spectroscopy \cite{Hayash52, Gross56,Nikiti69,Matsum96}. Because of the positive parity of the valence (symmetry $\Gamma_{7}^{+}$) and conduction (symmetry $\Gamma_{6}^{+}$) bands, the yellow series starts with the $\it{n}$=2 term. The 1s yellow series exciton level is split due to electron-hole exchange interaction into a forbidden singlet paraexciton ($\Gamma_{2}^{+}$ ; 2.021 eV at 2K) and a triply degenerate orthoexciton ($\Gamma_{5}^{+}$ ; 2.033 eV at 2 K). The spin configuration of paraexcitons is a purely triplet state and thus optical transitions to the ground state are strictly forbidden to all orders. Consequently, the 1s paraexciton has a very long lifetime and it has long been recognized as a promising candidate for the observation of exciton BEC \cite{Mysyro80, Fortin93}. The resonance energies of the $\it{n}\geq$2 terms obtained by linear absorption measurement can be well reproduced using the hydrogen formula, thus providing the basic parameters to characterize hydrogenic wave functions. The resonance of the 1s exciton state, however, has a significant deviation from the position extrapolated from the higher Rydberg lines ($\it{n}\geq$2). Such a deviation is known as the central cell correction of the 1s exciton state. It is caused by the non-parabolicity of the electron and hole bands at large wave vector due to the small Bohr radius of 1s excitons (comparable to the lattice constant \cite{Kavoul97}) and to dynamical screening. There still remain some ambiguities on the precise electronic structure of the 1s exciton even if this information is crucial to determine the stability of the exciton BEC phase.

\begin{figure}[tb]
\begin{center}
\includegraphics[width=50mm]{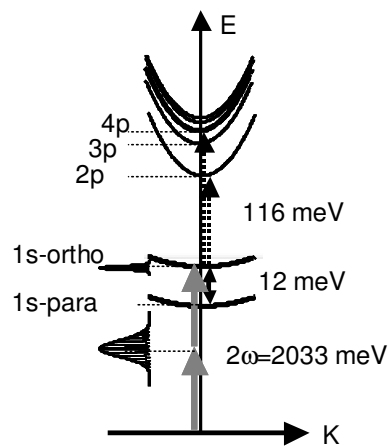}
\end{center}
\caption{Schematic energy diagram for the yellow series 1s and $\it{n}$p exciton states and their Lyman series transition in Cu$_{2}$O. }
\label{f1}
\end{figure}

In this paper, we report on the observation of the exciton Lyman series corresponding to transitions from a population of cold 1s orthoexcitons, created by a resonant two-photon process, to higher $p$ states up to the 5$^{th}$ term. 
Figure 1 schematically illustrates the time-resolved exciton Lyman absorption process occurring in our experiment along with an energy diagram of the yellow series excitons in Cu$_{2}$O. 
The split between ortho and paraexcitons is 12 meV. At a low density, the orthoexciton relaxes to the lower energy paraexciton state via a process involving the emission of transverse acoustic phonons, with a relaxation time of a few nanoseconds \cite{Jang04}. Recent experiments with a continuous wave probe light allowed to measure the Lyman signal associated with the 1s-2p transition of the paraexcitons. \cite{Jorger05, Karpin05} Time-resolved measurements with femtosecond pulses revealed the presence of both ortho and paraexcitons \cite{Gonoka04, Kubouc05}. 

A systematic analysis of the oscillator strength and line shape of the Lyman series allows to examine the detailed features of 1s excitons such as the wave function of the relative motion of electron-hole pair, the translational mass, and the distribution function of 1s excitons. For such an analysis, it is necessary to obtain first the 1s excitons with a narrow energy distribution. Since orthoexcitons can be generated by resonant two-photon excitation \cite{Frohli79}, we can apply a phase space compression scheme with ultrashort laser pulses, in analogy with the two-photon excitation of biexcitons in CuCl \cite{Gonoka02}. With this scheme, we can prepare orthoexcitons inside a very small phase-space, with an energy spread much smaller than the laser pulse bandwidth and the thermal energy of the lattice. In a one-beam excitation configuration, the orthoexcitons have an additional common kinetic energy of the the order of a few $\mu$eV which is gained from the finite photon momentum.

The details of the experimental setup and procedure have been described in previous papers. \cite{Gonoka04, Kubouc05, Gonoka05} A naturally grown single crystal has been cut in the form of a platelet shape with a 220 $\mu$m thickness along the (100) plane. The sample is cooled by contact with a copper block maintained at liquid Helium temperature. A tunable pump source with a wavelength centered around 1220 nm and a 500 fs pulse width (8 nm spectral width) is used. The pulse length is shaped from a 100 fs pulse by using a pair of gratings. The excitation pulse with 1 $\mu$J energy is focused onto a 3$\times$10$^{4}$ $\mu$m$^{2}$ area of the sample. The excitation light is linearly polarized and its orientation is set to maximize two-photon absorption. A mid-infrared tunable source with a wavelength centered around 10 $\mu$m is used for the weak probe pulse. The transmitted probe light is analyzed by a monochromator (0.1 meV spectral resolution) followed by a HgCdTe detector. The Cu$_{2}$O crystal shows a suitable transparency window between 112 and 137 meV, allowing the observation of transitions from the 1s orthoexciton to the $\it{n}$p states. 

\begin{figure}[tb]
\begin{center}
\includegraphics[width=80mm]{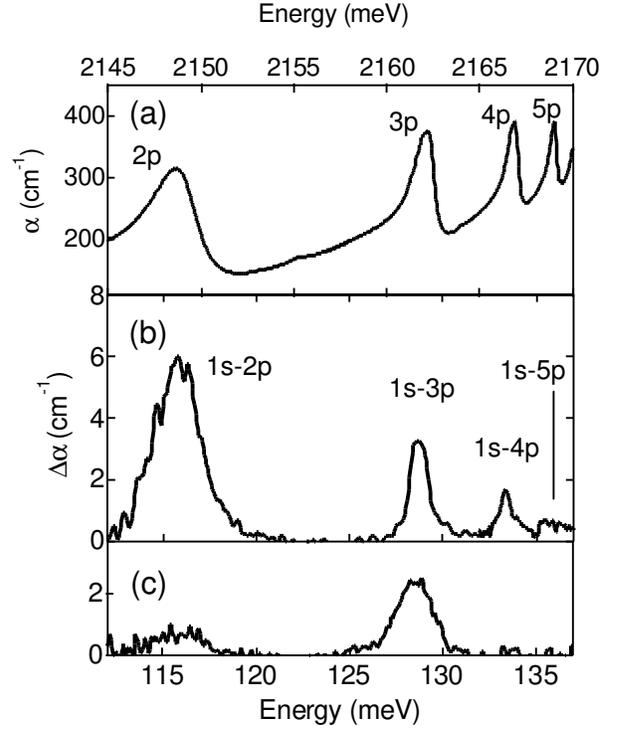}
\end{center}
\caption{Linear absorption spectrum (a). Induced absorption spectra of the 1s orthoexciton generated instantaneously by two-photon excitation (b), and the 1s paraexciton measured at 4 ns after the excitation (c).
}
\end{figure}

Figure 2(a) shows a linear absorption spectrum in the spectral region of $\it{n}$p orthoexciton states. In Figure 2(b), an induced absorption spectrum obtained under resonant two-photon excitation of 1s orthoexcitons at a low-density excitation (pump pulse energy 3 mJ/cm$^{2}$) measured after a pump-probe delay of 5 ps is shown. Under this excitation condition, a low exciton density was obtained, with a value estimated to be $n_{exciton}\sim$10$^{15}$ cm$^{-3}$ (See below). 
Four lines are observed and are positioned at 116, 129, 133, and 136 meV respectively. One can see that the spectral features obtained in Fig. 2(b) are in very good agreement with the ones shown in Fig. 2(a) if the energy shift corresponding to the 1s orthoexciton line $E_{1s ortho}$=2033 meV \cite{Uihlei81} is taken into account. Thus we can assign the series of Fig. 2(b) to the transition from the 1s to the $\it{n}$p orthoexcitons ($\it{n}$=2,3,4,5), i.e., the Lyman series. Their peak positions, widths, and area of absorption are summarized in Table 1.

It should be noted that the energy split between the 2p and 3p orthoexctions is close to that between the 1s ortho and paraexcitons. The 1s-2p paraexciton line should be carefully distinguished from the 1s-3p orthoexciton line as described in our previous paper\cite{Kubouc05}.
Figure 2(c) shows a transient absorption spectrum measured at a 4 ns pump-probe delay. 
The  1s-2p orthoexciton signal is weaker and the higher energy narrow lines are not present due to the decay of orthoexcitons into paraexcitons, while a relatively broad line appears around 128 meV. 
This new line is assigned to the 1s-2p transition of the paraexcitons.


\begin{table}[b]
\caption{Peak position, peak widths, and normalized integrated surfaces of the 1s-$\it{n}$p ($\it{n}$=2,3,4,5) lines. $\Delta E_{1s-np}$ is the energy estimated from the 1s and $\it{n}$p absorption lines.
}
\begin{tabular}{ccccc}
\hline
 & $E$ (meV) & $W$ (meV) & $S$ & $\Delta E_{1s-np}$\cite{Gross56,Uihlei81}\\
\hline
1s-2p & 115.9 & 3.0	&  1.0	& 115.8 \\
1s-3p & 128.8 & 1.1	&  0.26	& 129.0 \\
1s-4p & 133.4 & 0.7	&  0.11	& 133.9\\
1s-5p & 135.5 &    &  0.03	& 136.0\\
1s-2p(para) & 128.5 &  2.3	& & 127.6\\
\hline
\end{tabular}
\end{table}

The linewidth of the transitions of the exciton Lyman series has a systematic $\it{n}$-dependence. A similar behavior has been observed in $\it{n}$p absorption spectra and has been discussed in terms of an $\it{n}$-dependent LO phonon scattering rate from $\it{n}$p to 1s state\cite{Shindo74}. Figure 3 shows the linewidths of the 1s-$\it{n}$p transitions ($\circ$). The total width of the lines has several contributions: the energy distribution of 1s excitons, the intrinsic linewidths of the 1s state, and the width of the final $\it{n}$p states. The finite energy distribution of the 1s excitons, $f_{1s}$, contributes to the broadening because of the difference in the mass parameters between the 1s and $\it{n}$p excitons schematically shown in Fig. 1. The widths of the Lyman lines can be expressed as: ($\Delta$E$_{1s-np}$)$^{2}\sim$($\Delta$E$_{np}$)$^{2}$+((m$_{1s}$/m$_{2p}$-1)$f_{1s}$)$^{2}$. The intrinsic linewidth of the 1s exciton state is much narrower ($\gamma_{1s}\ll$0.1 meV) than the spectral resolution of the present experiment and is neglected in our analysis. 
We estimate the linewidth of the $\it{n}$p states by fitting the linear absorption data lines with asymmetric Lorentzian curves following the model of Toyozawa \cite{Toyoza64}. The obtained result is also plotted in Fig. 3 ($\bigtriangleup$). From the difference of the width between the 1s-4p and the 4p lines ($<$0.1 meV), the upper limit of the energy spread of initial 1s orthoexcitons, $f_{1s}$, is estimated to be 0.36 meV/(3m$_{e}$/1.7m$_{e}$-1)$\sim$0.5 meV, which is narrower than the energy spread expected for the 4.2 K thermal equilibrium exciton gas, 0.7 meV. This indicates that the initial super cooled 1s orthoexcitons are still at a temperature below that of the lattice even at a delay of 5 ps.

\begin{figure}[tb]
\begin{center}
\includegraphics[width=40mm]{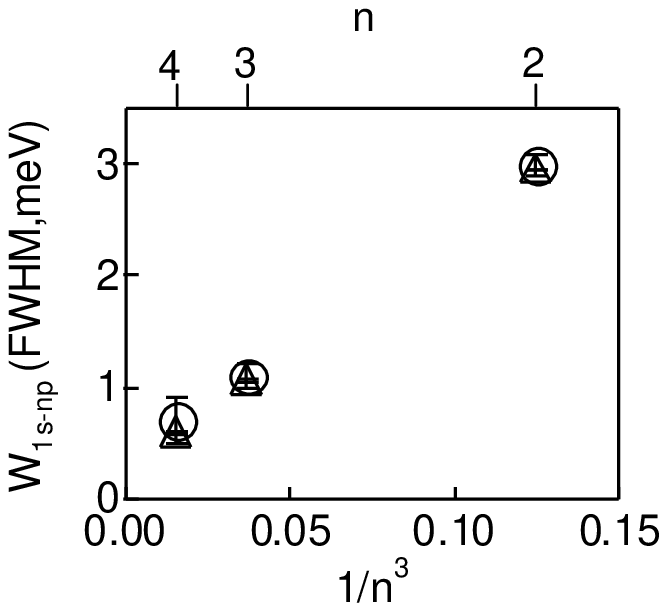}
\end{center}
\caption{Linewidth of 1s-$\it{n}$p lines ($\circ$) and $\it{n}$p absorption lines ($\bigtriangleup$). }
\end{figure}

The wave functions of $\it{n}$p states are well described by a hydrogenic model, since the spectral positions of the $\it{n}$p exciton states ($\it{n}\geq$2) are well described with a Rydberg series $E_{n}$=17525-786/$\it{n}$$^{2}$ cm$^{-1}$ \cite{Nikiti69}.  However, the 1s exciton state shows a significant deviation from this Rydberg series which has been discussed by several authors. Most recently, Artoni et al. analyzed the AC Stark effect under resonant pumping of the 1s-2p transition of orthoexciton and used a value $\mu_{1s-2p}$=6.3 e$\AA$ (30 D) in direct analogy with hydrogen atoms \cite{Artoni02}. Kavoulakis et al. examined the shrinkage of the Bohr radius of the 1s exciton state caused by the central cell correction due to the non-parabolicity of the valence and conduction bands.\cite{Kavoul97} They took into account the quadratic terms of the wave vector dependence of the bands under the constraint that the known value of translational mass of 1s orthoexciton, 3m$_{e}$ must be reproduced. This analysis gave a Bohr radius $a_{1s}$ =5.3 $\AA$ and a binding energy E=164.9 meV. They found that the wave function of the 1s exciton is well approximated with such a modified 1s function \cite{Uihlei81}. Recently, it was pointed out that such a small effective Bohr radius of the 1s excitons reduces the overlap between the 1s and 2p exciton wave functions, yielding a revised dipole moment of 1.64 e$\AA$ \cite{Jorger05A}. 
The systematic analysis of the exciton Lyman absorption provides a unique opportunity to determine the "unknown" 1s exciton wave function from a set of projections to the known $\it{n}$p states through dipole matrix element $\mu_{1s-np}=e<u_{1s}|z|u_{np}>$. 

The integrated area of the exciton Lyman absorption line 1s-$\it{n}$p can be expressed as follows: 

\begin{center}
$\int_{1s-np}\Delta\alpha(E)dE=\frac{n_{1s}4\pi^{2}E_{1s-np}|\mu_{1s-np}|^{2}}{\hbar c\sqrt{\varepsilon}}$ (1)
\end{center}
where $E_{1s-np}$, $\mu_{1s-np}$ and $\varepsilon$ are the 1s-$\it{n}$p transition energy, dipole moment and background dielectric constant at the probe frequency respectively. We determine the background dielectric constant $\varepsilon\sim$6 by measuring the interference fringe pattern of the absorption spectrum in the mid-infrared region. Knowing the density of the 1s exciton and from the measured integrated absorption $S_{1s-np}=\int_{1s-np}\Delta\alpha(E)dE$, we can determine the dipole moment $\mu_{1s-np}$. In practice, however, it is difficult to estimate the absolute density of the 1s exciton from the absorbed photon number, which strictly depends on the detailed pumping condition especially in the case of a two-photon excitation with short pulses. Therefore in order to determine the wave function of the 1s exciton state, we rely on the relative intensity of the integrated absorption of the 1s-$\it{n}$p lines $S_{1s-np}/S_{1s-2p}$, which are density independent and insensitive to the pump condition. Indeed, in the low-density excitation limit, Eq.(1) yields the ratio, $S_{1s-np}/S_{1s-2p}=\mid\mu_{1s-np}/\mu_{1s-2p}\mid^{2}\times(E_{1s-np}/E_{1s-2p})$.

\begin{figure}[tb]
\begin{center}
\includegraphics[width=80mm]{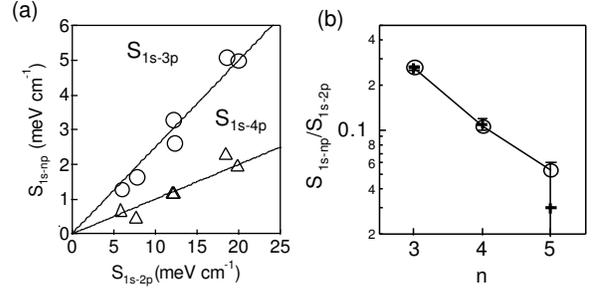}
\end{center}
\caption{(a) Integrated spectral area of 1s-3p lines ($\circ$) and 1s-4p lines ($\bigtriangleup$) as a function of area of the 1s-2p lines.
(b) Integrated area of 1s-$\it{n}$p induced absorption lines normalized by that of 1s-2p line. +: Experimental results. $\circ$: Best fit (4.2 e$\AA$). 
}
\end{figure}

Figure 4(a) shows the integrated spectral area of the 1s-$\it{n}$p induced absorption ($\it{n}$=3,4) as a function of the area of the 1s-2p transition. 
By varying the excitation intensity, we find that the integrated areas of the 1s-3p and 1s-4p transitions are proportional to the integrated area of the 1s-2p line, i.e. the 1s exciton density. 
From the fitting of the result shown in Fig. 4(a), we obtain the ratio $S_{1s-3p}/S_{1s-2p}$=0.26$\pm$0.01 and $S_{1s-4p}/S_{1s-2p}$=0.11$\pm$0.01. Figure 4 (b) shows the S$_{1s-np}$/S$_{1s-2p}$ integrated area of the 1s-$\it{n}$p lines ($\it{n}$=3,4,5) normalized by that of the 1s-2p line. The crosses show the experimental results. With our analysis, we obtain $S_{1s-np}/S_{1s-2p}=|\mu_{1s-np}/\mu_{1s-2p}|^{2}\times(E_{1s-np}/E_{1s-2p})$ where $\mu_{1s-np}=e<u_{1s}|z|u_{np}>$ and $E_{1s-np}$ are the transition dipole moment and resonance energy of 1s-$\it{n}$p transition respectively.
From the observed value of $S_{1s-np}/S_{1s-2p}$, we obtain a best fit value of the 1s Bohr radius of $a_{1s}$=7.9 $\AA$, which gives a dipole moment of $\mu_{1s-2p}$=4.2 e$\AA$. 
It should be noted that the 1s Bohr radius extracted from the positions of the $\it{n}$p exciton series ($\it{n}$$\geq$2) is $a_{1s}$=11.1 $\AA$. 
According to Eq. (1), the spectrally integrated absorption area $S_{1s-2p}\sim$1 meVcm$^{-1}$ corresponds to an exciton density $n_{1s}\sim$6$\times$10$^{13}$ cm$^{-3}$ under our relatively low excitation conditions. 

Equation (1) implies that we can determine the value of the transition dipole moment from the absorption strength if we know the density of excitons. To cross check the validity of our procedure to determine the dipole moment and Bohr radius based on the ratio $S_{1s-np}/S_{1s-2p}$, we also measured the induced Lyman absorption under one photon excitation at a 600 nm wavelength, corresponding to the phonon side band of the 1s orthoexciton, where we can expect a quantum efficiency close to 100 $\%$ for photon-exciton conversion. With an absorbed photon density of 2$\times$10$^{15}$ photons/cm$^{3}$, we measured an induced absorption signal at a 50 ps delay $S\sim$53 cm$^{-1}$meV in the spectral region from 112 meV to 137 meV, which includes the contributions from the 1s-$\it{n}$p orthoexcitons and the 1s-2p paraexcitons. At such short delay time, we can neglect the decay of the orthoexcitons. We obtain $\mu_{1s-2p}$ of 4.8 e$\AA$. The value is slightly overestimated since the contribution of the 1s-3p transition is neglected. Thus the obtained value is consistent with the estimation from the $S_{1s-np}/S_{1s-2p}$ ratio.

With a hydrogenic model, the main cause for the deviation of the 1s exciton level from the Rydberg series is the dynamic effect on the electron-hole screening. Here, we introduce the AC dielectric constant at a frequency close to the exciton binding energy. The measured dielectric constant is $\varepsilon\sim$6. The effective Bohr radius is determined from the binding energy $E_{bin}$ and dielectric constant $\varepsilon$ with  $a_{B}=2\pi e^{2}/E_{bin}\varepsilon\epsilon_{0}$. 
$E_{bin}\sim$150.8 meV and $\varepsilon\sim$6.0 in turn give a value $a_{B}\sim$7.8 $\AA$, which agrees with our estimate. This indicates that the hydrogenic model with only a correction due to screening by the AC dielectric constant is suitable to explain the general feature of 1s excitons in Cu$_{2}$O. 

In summary, we observed the Lyman series up to the 5p term of the yellow series exciton in Cu$_{2}$O using femtosecond time-resolved mid-infrared spectroscopy. From the measured narrow linewidths of the higher Rydberg series, we inferred that the 1s orthoexcitons have an effective temperature lower than the lattice temperature even 5 ps after their creation by direct two photon excitation. This confirms that phase space compression of the resonant two-photon excitation with femtosecond pulses is an effective method to create a super cooled exciton gas. From a systematic analysis of the Lyman series absorption strength, we obtain a Bohr radius of the 1s exciton $a_{1s}$ =7.9 $\AA$ and a 1s-2p transition dipole moment $\mu_{1s-2p}$=4.2 e$\AA$. 
The obtained dipole moment enables us to determine the exciton density directly from the spectrally integrated absorption area; $n_{1s}$/$S_{1s-2p}$=6$\times$10$^{13}$ cm$^{-3}$/meVcm$^{-1}$. 
This novel method for the quantitative measurements of exciton density is important for the study of high-density effects in excitons including exciton Bose-Einstein condensation, Auger recombination and exciton Mott transition.

\section*{Acknowledgment}
The authors are grateful to R. Shimano, M. Kubouchi, N. Naka, and K. Yoshioka for stimulating discussions and support during the experiments. We also thank to J. B. H\'{e}roux for critical reading the manuscript. This work is supported in part by KAKENHI (S) from JSPS.

\end{document}